# A Wave Trapping Dispersive Delay Structure Based on Substrate Integrated Waveguide (SIW)

Hossein Nazemi-Rafi [1], Masoud Movahhedi [2*]

[1,2] Electrical Engineering Department, Yazd University, Yazd, Iran
[*] movahhedi@yazd.ac.ir

**Abstract:** A dispersive delay structure (DDS) is a device that creates a specified group delay function over frequency bandwidth. In this paper, the idea of trapping the electromagnetic waves in a feedback loop is proposed to achieve higher group delay swing. The proposed structure includes two transmission lines and two junctions that act as power divider/combiner. The transmission lines can be either dispersive or non-dispersive. The group delay in the proposed configuration is proportional to the injected power into the feedback loop. So, an increase in the group delay swing can be done by improving the junctions or by adding more feedback loops to the structure. Based on this configuration, a novel DDS, using the substrate integrated waveguide (SIW) transmission line, is introduced. According to the experimental results, the proposed device with a relatively compact size creates high group delay swings (*9 ns*), and its insertion loss is about *7 dB*

## 1. Introduction

High-cost and high-power use are some of the issues with digital signal processing (DSP) at high frequencies. Using real-time analog signal processing (ASP) systems is a solution to these problems [1]. One of the fundamental analog components in some ASP systems is a dispersive delay structure (DDS), which provides group delay difference over a frequency bandwidth. An ideal two-port DDS has a transfer function of $e^{j\phi(\omega)}$, and its group delay, i.e., $\tau(\omega) = -d\phi/d\omega$, is a function of the frequency [1]. Therefore, the main characteristic of any DDS is its group delay, which determines the bandwidth of the device. The scattering parameters are of secondary importance, but it is suitable to have a DDS with low insertion and high return loss. From the group delay point of view, filters and DDSs are in opposite of each other, because dispersion is not desirable in filters [2], as it is the whole point of the DDS application.

In general, there are two types of DDS: reflection-type and transmission-type [1]. Reflection-type DDS is a one-port network that needs an additional device, like a circulator, for the creation of the second port [3]. The transmission-type DDS is a simple two-port network and does not require any additional tools. But, for the same structure size, the reflection-type DDS creates higher group delay value (for example, a short cutted waveguide compared to a simple two-port waveguide transmission line).

Both types of DDS have the same goal, which is to create a specific group delay function. For example, group delay with step function could be used in a system of analog signal processing with the purpose of frequency separation [1]. With this technique, it is possible to detect that the frequency channel is occupied or not. However, one of the main applications of DDS is pulse compression in radar systems. In this case, the group delay needs to be linear (or semi-linear) function of frequency [1]. In general, the creation of such group delay functions is not possible by using a single DDS, and it is necessary to combine different group delay functions to obtain the desired form. This could be done by cascading uncoupled [4] or coupled [5] DDSs. Thus, the first step is designing a single DDS as a unit cell.

Generally, there are three approaches for designing a DDS: using dispersive TLs (or media), dispersive configurations, and combination of both. C-section structure is one of the most popular configurations for a DDS, which has been used for decades (as meander lines) and recently was studied by Gupta et al. [4, 5]. Configuration for a single C-section is a simple backward directional coupler which is short-ended at one end [5]. The directional coupler could also be formed to a feedback loop (ring) by connecting ports of one of the arms of the coupler [6]. Coupled-line all-pass phasers consists of $N$ coupled sections as a longitudinal cascade of commensurate coupled transmission lines which are based on cascaded C-sections have been proposed in [7]. These structures exhibit $N$ group-delay peaks within a harmonic frequency band. Subsequently, step-discontinuity and continuously nonuniform C-sections which are generalized form of the coupled-line all-pass phasers and provide enhanced bandwidth and diversity compared to conventional C-sections were introduced [8].

Dispersion in these configurations could be achieved even by using non-dispersive TLs like microstrip. However, passing wave through a single dispersive TL like a waveguide [9] or a composite right/left-handed (CRLH) line [10, 11] is a practical and straightforward way to obtain a dispersion. Combining these two methods (dispersive configuration and TLs) could lead to a compact DDS and a high group delay value [12, 13].

SIW is a printed rectangular waveguide and, like other dispersive TLs, could be used solely or as a construction block for the creation of dispersive structures. Recently, half-mode SIW (HMSIW) TL [14] has been used for the realization of microwave coupled ring phasers [6]. Loss is a significant issue in designing the DDSs, especially at group delay peaks [14]. This problem could be solved by inserting an amplifier into a single DDS [14, 15], or between DDSs in the cascading configuration. The second problem is the size of the cascaded structure, which could be solved by inserting DDS into a feedback loop configuration [16, 17]. In these active devices, the input signal repeatedly goes through a dispersive structure, and a switch controls the number of repetitions. The feedback loops, also, includes an amplifier to compensate iteration loss.

In this paper, a new configuration of dispersive delay structures based on the wave trapping or feedback mechanism is proposed. The proposed configuration, that is called the wave trapper, consists of transmission lines and two power dividers/combiners to create a feedback loop. The power dividers inject the power into the feedback loop, and it will be shown that the group delay is directly related to the trapped energy. This power can be increased by using power dividers with higher dividing ratio, or by adding more loops to the structure. There are no restrictions on the type of used transmission lines; but using dispersive TLs leads to the higher group delay peaks. Moreover, for the first time, we have used SIW TLs to implement the proposed configuration of DDS, which not only are dispersive near the cut-off frequency, but they are also easy to use in multilayer structures. In other words, the implementation of the wave trapper is realizable and easy by using SIW transmission lines, which also removes radiation losses compared to HMSIW [14]. Also, using SIW TLs allows adding more feedback loops to the wave trapper configuration, which increases the group delay swings. By decreasing substrate losses, extremely high group delay, in a multi-layer SIW wave trapper, is achievable.

This paper is organized as follows: Section 2 describes SIW based DDS. Section 3 presents a novel configuration of a wave-trapper and its group delay mathematical function. Sections 4 and 5 presents the realization of the wave-trapper by SIW TLs and the test results, respectively. Finally, Section 6 illustrates conclusions.

## 2. SIW Based DDS

Conventional waveguide TLs, similar to CRLH TLs [11], display dispersion characteristic due to their non-linear phase constant [16]:

$$\beta = \sqrt{k^2 - k_c^2} \quad (1)$$

where $k = \omega\sqrt{\mu\epsilon}$ is the wavenumber at frequency $\omega$, and $k_c = \pi/a$ is the wavenumber at the cut-off frequency $\omega_c$ (assuming $TE_{10}$ mode in a rectangular waveguide with a width of $a$). Waveguides have good dispersive traits near their cut-off frequencies, because with increasing the frequency far from the cut-off frequency, the phase constant becomes linear ($\beta \approx k$), and therefore, the group delay becomes constant.

In 1967, Bromley and Callan used 91.5 meters of short-circuited WG11A ($a = 2.29\ in$, $b = 1.145\ in$ and $f_c = 2.577\ GHz$) and a circulator to create a two-port DDS with effective length of 183 (2×91.5) meters [9]. Because of the long length of the device, the DDS had been compacted in a 20-feet container. The operation frequency band and the group delay difference of the DDS was *2.65 GHz - 2.85 GHz* and *1.05 µs*, respectively. In other words, WG11A waveguide without any change was used as a DDS near its cut-off frequency. This rectangular waveguide can also be used as a non-dispersive TL with only $TE_{10}$ mode propagating within *3.3 GHz - 4.9 GHz* bandwidth. Fabrication of rectangular waveguide by a printed circuit board (PCB) technology became possible by introducing substrate integrated waveguide (SIW) [19]. Recently, SIW transmission lines are used by many engineers in the design of microwave and antenna devices [18]. Similar to the waveguide, SIW TL can also be used as a dispersive structure near its cut-off frequency. By passing the electromagnetic wave through a SIW DDS (Fig. 1(a)), the wave would be gained a delay of:

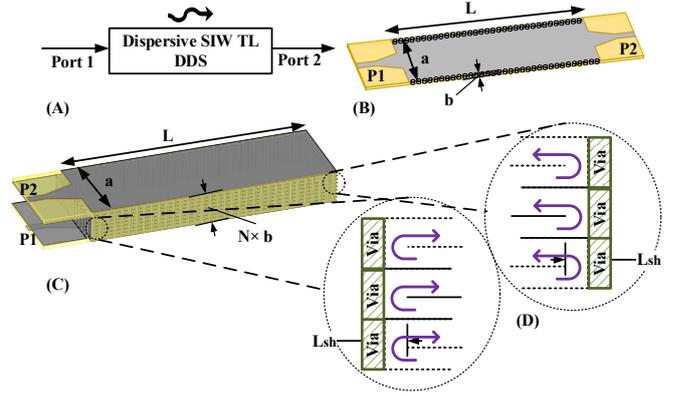

***Fig. 1. (a)*** *Schematic for SIW TL based DDS.* ***(b)*** *SIW TL with the length of L in a one-layer substrate.* ***(c)*** *SIW TL with length of N×L in N-layer substrate.* ***(d)*** *Wave transmission mechanism from one layer to the next by short-cut aperture. (a: width of SIW TL, b: substrate thickness).*

$$\tau = -d(-\beta L)/d\omega = k\sqrt{\mu\epsilon}L/\sqrt{k^2 - k_c^2} \quad (2)$$

where $L$ is the length of the SIW TL (Fig. 1(b)). The group delay $\tau$ can be increased, simply, by cascading DDSs, which in this case is equal to increasing the length $L$. To compact a long and straight SIW TL in a single-layer substrate, it can be stacked up in a multi-layer structure (Fig. 1(c)). The wave is guided to the next layer by a very short (≈10 mil) aperture, as seen in Fig. 1(d).

The scattering parameters and the group delay for the structure of Fig. 1 with physical parameters of *a = 0.7 in, b = 32 mil, εr=3.38, L=2 in, Lsh=10 mil, Dvia=20 mil, Pvia=2D* and *N = 8* are shown in Fig. 2. The cut-off frequency of SIW TL is 4.6 GHz. The upper and the lower frequencies of the DDS bandwidth are *4.7 GHz* and *5.3 GHz*, respectively, and the group delay difference *Δτ* is *6 ns (11 ns – 5 ns)*. At higher frequencies, the group delay is constant (about 4 ns), and SIW TL becomes nearly a non-dispersive TL for propagating $TE_{10}$ mode until the next cut-off frequency (in this case $TE_{20}$). The minor issue with DDS that operates near its cut-off region is that the group delay peak occurs at cut-off frequency, so, the full potential of the DDS could not be reached (note that, the starting frequency of the band is considered to be a cut-off frequency, too). But the major issue is that these DDSs could not be used in dispersion engineering to create arbitrary group delay functions. Nevertheless, these kinds of DDSs are practical and useful for semi-linear and high group delay applications. But it is suitable to design a DDS with its group delay peak within the pass-band of the structure (bell-form).

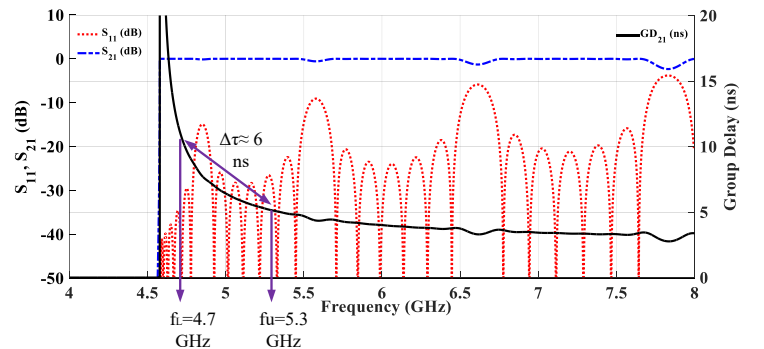

***Fig. 2.*** *Scattering parameters and the group delay for two-port SIW based DDS illustrated in Fig. 1(c).*

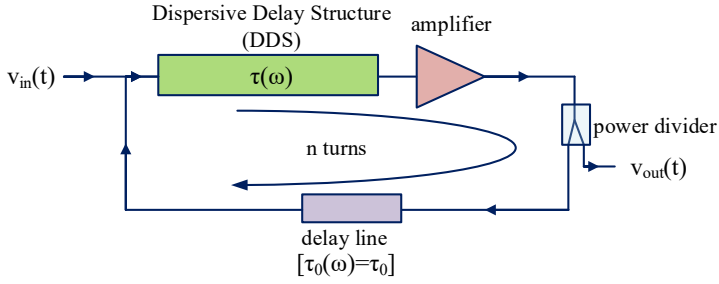

*Fig. 3. Configuration of the feedback loop proposed by Nikfal et al. to increase the dispersion and group delay [17].*

To reach the higher values of the group delay, a dispersive structure can be used as a unit cell in a cascaded configuration. The cascading method is simple but creates two issues. The first problem is attenuation, especially in the case of substrate-filled SIW compared to the air-filled waveguide. The loss could be compensated by adding amplifiers between DDSs. However, the second problem still remains unresolved, which is the long total size of the structure. For overcoming the cascading problem, the idea of using a feedback loop, to increase the group delay, was introduced by Fetisov et al. [16] in 1998 and Nikfal et al. in 2011 [17]. The configuration of their active DDS is shown in Fig. 3. By this technique, the input signal $V_{in}(t)$ goes through the DDS multiple times, and the final group delay difference would be multiplied. A similar feedback loop was used by Gupta et al. in 2015 [6] and Emara and Gupta in 2019 [14], by converting directional coupler to a feedback ring.

## 3. Proposed DDS Based on Wave Trapper

In this paper, a new configuration for a DDS based on the wave trapping phenomena is proposed. Unlike other feedback loops [6, 14], this configuration does not need edge coupling to inject power to the feedback loop. The DDS is a wave trapper based on non-dispersive and dispersive TLs without any active devices. The proposed configuration is realized by SIW TLs to show that this structure could be easily implemented.

Fig 4. shows the schematic for the wave-trapper, which consists of two TLs with lengths of $L_1$ and $L_2$ and two junctions in the form of γ, as ideal power combiner/dividers. Let us assume an input wave $E_1$ as the system input. At the first γ-junction, the input wave combines with the wave from port $P_2$ ($E_2$). So, wave at the output of the first γ-junction is:

$$E_3 = E_1 + E_2 \quad (3)$$

Then, $E_3$ goes through a TL with the length of $L_1$, and experiences a phase shift of $\theta_1=\beta L_1$. Thus, wave at the input of the second γ-junction is $E_4=E_3 e^{-j\theta_1}$. The second γ-junction acts as an ideal power-divider, and splits the wave to:

$$E_5 = kE_4, \quad E_6 = \sqrt{1-k^2}E_4 \quad (4)$$

where $k^2/(1-k^2)$ is the power dividing ratio. $E_6$ goes to the output, and wave $E_5$ experiences a phase shift of $\theta_2=\beta L_2$ and arrives at port $P_2$ as $E_2= E_5 e^{-j\theta_2}$. Therefore, some portion of the power is trapped in the feedback loop, and the group delay of the structure could be extremely high. To validate this theory, it would be necessary to find a mathematical formula for the group delay in the wave-trapper system. Thus, the relation between the output and the input waves ($E_6$ and $E_1$) should be obtained. After simple mathematical steps, we have:

$$E_6 = E_1 \sqrt{1-k^2} e^{-j\theta_1} / (1 - k e^{-j\theta_t}) \quad (5)$$

where $\theta_t = \theta_1 + \theta_2$. So, the phase difference between the output and the input waves is:

$$\begin{aligned}\phi_{21} &= -\theta_1 - \theta_{loop} \\ \phi_{21} &= -\theta_1 - \tan^{-1}(k \sin\theta_t/(1-k\cos\theta_t))\end{aligned} \quad (6)$$

The group delay can be obtained by $\tau_{21} = -d\phi_{21}/d\omega$:

$$\begin{aligned}\tau_{21} &= L_1\, d\beta/d\omega \\ &+ (L_1 + L_2)\, d\beta/d\omega \\ &\times (k\cos\theta_t - k^2)/(1 + k^2 - 2k\cos\theta_t)\end{aligned} \quad (7)$$

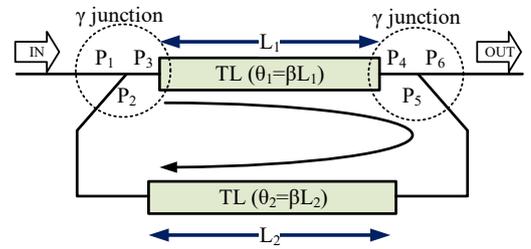

*Fig. 4. Proposed wave-trapper configuration.*

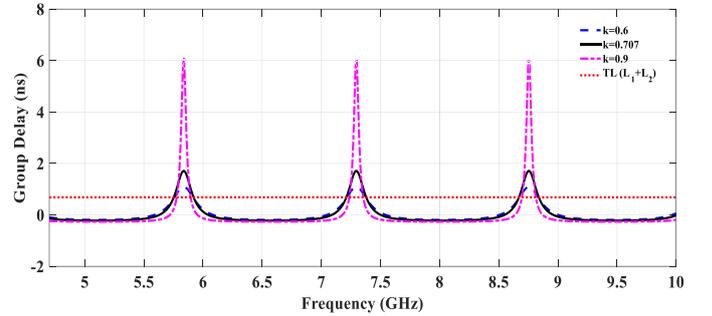

*Fig. 5. The Group delay of the wave-trapper (based on non-dispersive TLs) with different power-dividing ratios compared to a single TL (doted-line) ($L_1$=0.4 in, $L_2$=4 in and $\varepsilon_r$=3.38).*

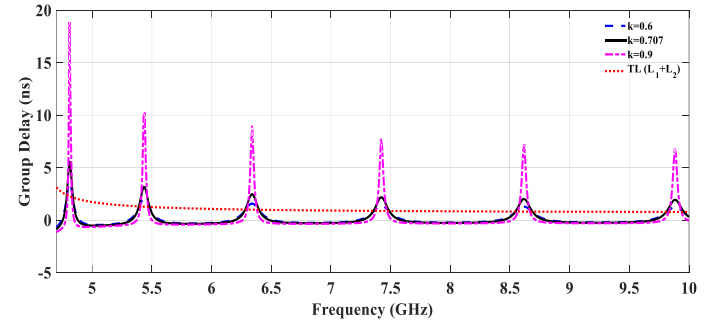

*Fig. 6. The Group delay of the wave-trapper (based on dispersive SIW TLs) with different power-dividing ratios compared to a single SIW TL (doted-line) ($L_1$=0.4 in, $L_2$=4 in, $\varepsilon_r$=3.38 and $f_c$=4.6 GHz).*

Fig. 5 shows the group delay of a wave-trapper based on non-dispersive TLs. The doted-line shows the group delay of a non-dispersive simple line with the length of $L_1+L_2$.

As seen in this figure, the group delay is a periodic function of the frequency, and its maximum occurs at $\theta_t = 2m\pi$. The group delay difference ($\Delta\tau = \tau_{max} - \tau_{min}$) is proportional to parameter $k$ (Fig. 5). With $k = 0.707$, when the power dividing ratio is 1:1 (half of the power is trapped into the loop), the group delay difference is about *2 ns*. If k increases to 0.9 (catching 80 percent of the energy into the loop with 4:1 power divider), $\Delta\tau$ increases to *6 ns*.

The group delay of a wave-trapper with SIW TL as a dispersive line is illustrated in Fig. 6. The cut-off frequency of SIW TL is *4.6 GHz*, and near this frequency, the TLs in the wave-trapper system are dispersive. In this dispersive region, the group delay peaks are much higher, and closer to each other (Fig. 6.). For example, the group delay difference near *5.5 GHz* (for $k = 0.9$) is *10 ns*, which is *4 ns* higher than the wave-trapper based on non-dispersive TLs.

For the realization of the proposed configuration, the γ-junction should act as a good power divider/combiner. In the next section, the wave-trapper would be realized by SIW TLs.

## 4. SIW Based Wave Trapper

Fig. 7 shows the top view of the proposed SIW wave-trapper. The width of all SIW TLs is *a=0.7 in*, the height of the substrate is *b = 32 mil*, and the substrate dielectric constant is $\varepsilon_r = 3.38$ (Rogers RO4003). The feedback loop is an elliptical ring with inner radiuses of $R_1 = 0.68$ *in* and $R_2 = 0.36$ *in*.

Scattering parameters and the group delay of the wave-trapper are given in Fig. 8. There are four group delay peaks between *5 GHz* to *8 GHz*, but only three of them have suitable return (insertion) loss. The two peak frequencies of $f_2 = 6.6$ *GHz* and $f_3 = 7.4$ *GHz* are in the middle of the pass-band, but $f_1 = 5.8$ *GHz* is near a cut-off frequency. This DDS creates a maximum group delay difference of *2.5 ns*, as shown in Fig. 8.

To visualize the trapped waves in the feedback loop, the graphic plots of the electric field intensity in the wave-trapper are shown, for $f_2 = 6.6$ *GHz* and $f_3 = 7.4$ *GHz*, in Fig. 9. The ratio of the electric field amplitude in the loop to the electric field in the straight line, for $f_2 = 6.6$ *GHz*, is about *1.44*; so, the power-dividing ratio is:

$$k^2/(1-k^2) \approx 1.44^2 = 2.1 \qquad (8)$$

Therefore, 63 percent of the energy is trapped in the loop. The electric field amplitude ratio is about *1.69* for $f_3 = 7.4$ *GHz*, and 74 percent of the energy is rotating in the ring (Fig. 9).

But, the group delay of the designed structure in Fig. 7, as seen in Fig. 8, is slightly different from the ideal group delay that has been illustrated in Fig. 6, due to non-ideal power divider/combiner. The γ-junctions can be improved by adding a metallic blade to each junction (for SIW TL, adding metallic vias). Fig. 10 shows the structure of the improved wave-trapper. Scattering parameters and the group delay of the improved wave-trapper are given in Fig. 11. The group delay function is more similar to Fig. 6. The three peak frequencies are in the middle of the pass-band, but the return loss for $f_1 = 5.8$ *GHz* is not as good as the other two

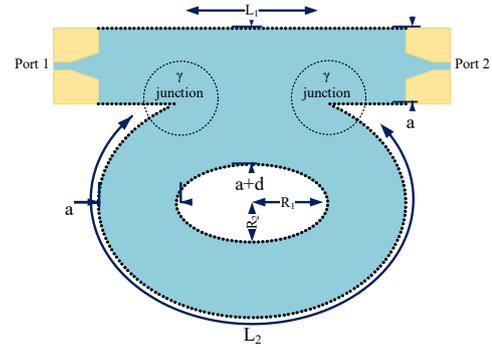

**Fig. 7.** *Top view of the proposed DDS based on SIW based wave-trapper (d=0.56 in).*

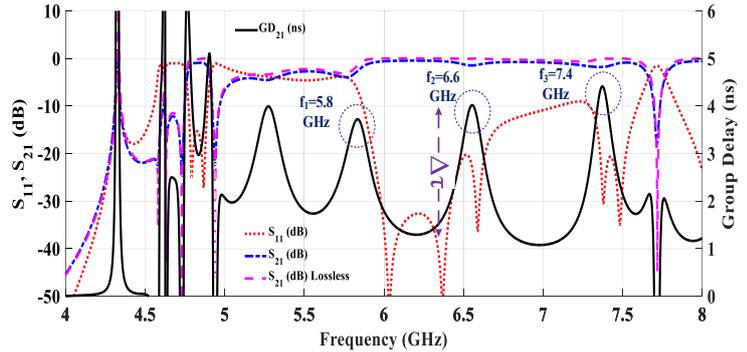

**Fig. 8.** *Scattering parameters and the group delay of the wave trapper based DDS in Fig. 7 (CST was used for simulation).*

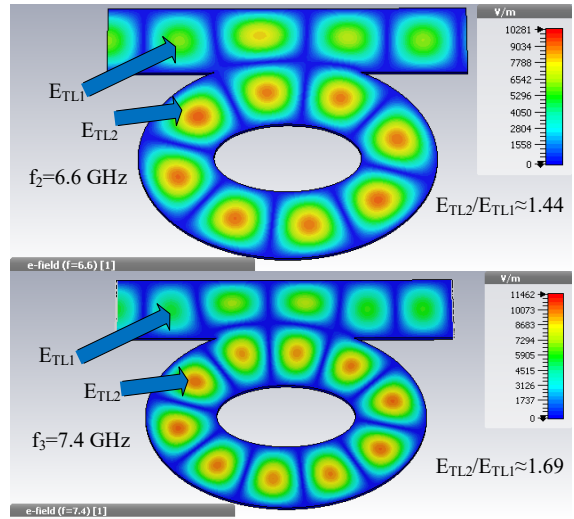

**Fig. 9.** *Visualization of the electric field in the wave-trapper at two peak frequencies ($f_2$=6.6 GHz and $f_3$=7.4 GHz).*

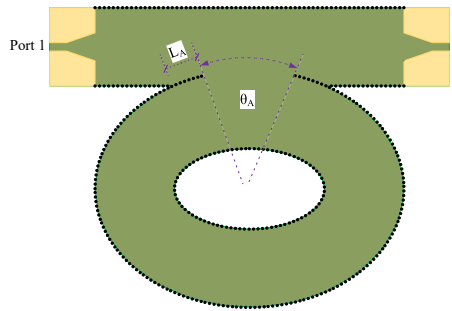

**Fig. 10.** *Configuration of the improved wave-trapper by adding metallic blades to the junctions (a=0.7 in, b=32 mil, $\varepsilon_r$=3.38, $R_1$=0.66 in, $R_2$=0.36 in, d=0.56 in, $\theta_A$=34° and $L_A$=0.29 in).*

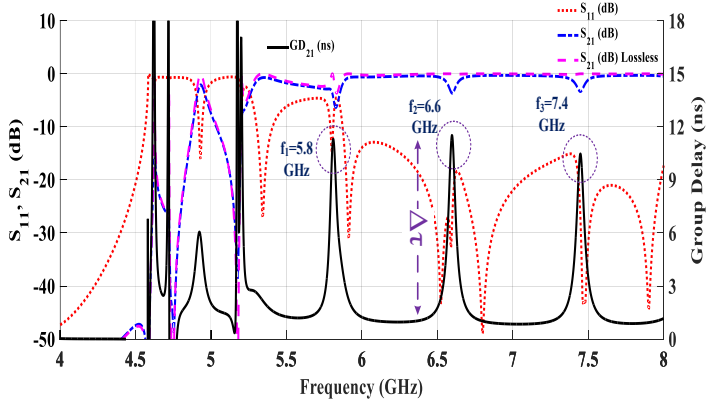

*Fig. 11. Scattering parameters and the group delay of the wave-trapper of Fig. 10.*

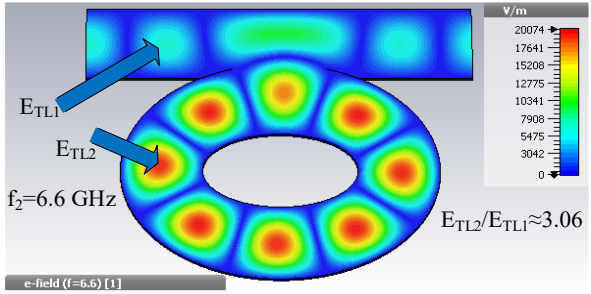

*Fig. 12. Visualization of the electric field in the improved wave-trapper at the peak frequency of $f_2$ = 6.6 GHz.*

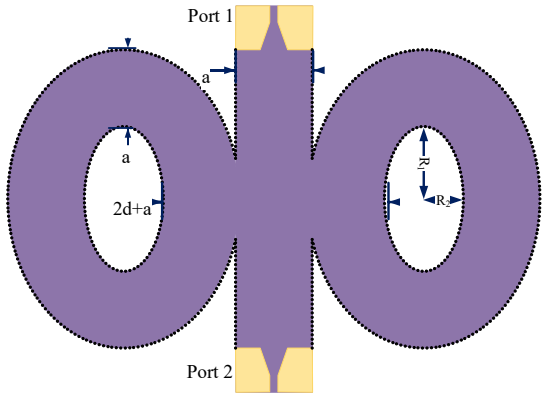

*Fig. 13. Configuration of a symmetric wave-trapper with two feedback loops (a=0.7 in, b=32 mil, $\varepsilon_r$=3.38, $R_1$=0.66 in, $R_2$=0.36 in and d=0.66 in).*

frequencies. The group delay difference Δτ is about *10 ns*, which is *7.5 ns* higher than the wave-trapper without metallic blades (Fig. 11). The electric field visualization at $f_2$ = 6.6 GHz is displayed in Fig. 12. The power-dividing ratio is about *9.4:1* and 90% of the energy is captured in the loop.

Configuration of the wave-trapper could be improved in many ways. The simple one is adding an extra loop in the system. Fig. 13 shows a symmetric wave-trapper with two loops. Scattering parameters and the group delay of the double-loop wave-trapper are shown in Fig. 14. The group delay difference Δτ is about *15 ns* (as seen in Fig. 14).

Adding more loops to the system increases the group delay of the structure. A symmetric wave-trapper with four feedback loops is shown in Fig. 15, and its loss and group delay characteristics are indicated in Fig. 16. Peak frequencies of the two pair loops are not matched, and there is a *200 MHz* frequency shift. By changing design parameters, the double peak could be reduced to a single high peak of *50 ns*, but in exchange for the peak being at the start of the pass-band (at the cut-off frequency).

The final step is cascading, which could be done in one-layer or multi-layer or a combination of both. But, the loss of the structure, especially at peak frequency, becomes high. Fig. 17 shows a 3D view of the structure. The size of the device is *6.4×2.7×0.2 $in^3$* (*162×68×5 $mm^3$*) which consists of 12 (2×6) cells. Return loss, insertion loss, and the group delay for peak frequency of *f = 6.6 GHz*, are shown in Fig. 18. This device could create *120 ns* group delay swing. But the insertion loss is about 3 dB, even by using low loss substrate (tan $\delta \approx 10^{-4}$). Insertion loss could be increased to 40 dB by using a normal substrate with medium loss (tan $\delta \approx 10^{-3}$).

It is important to note that the proposed structure presents a main disadvantage which is the losses at frequencies were delay is maximized. This appears to be inherent to the structure due to the wave trapping phenomena. This drawback of the structure, i.e., the loss problem, can be solved by using very low loss substrates or air-filled SIW (AFSIW) [21]. In addition, the use of an amplifier can be helpful to overcome this disadvantage as suggested in [14]. Besides the dielectric loss of the substrate in AFSIW, other losses should be considered such as: the transition from microstrip (or CPW) to SIW, the transition from SIW to AFSIW, leaking waves and conductor loss. Assuming the multi-layer AFSIW technology, creating extremely high group delay difference in a compact size is possible.

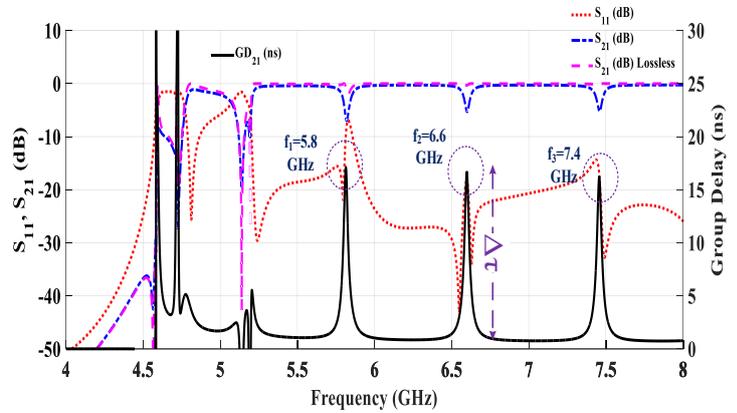

*Fig. 14. Scattering parameters and the group delay of the wave-trapper of Fig. 13.*

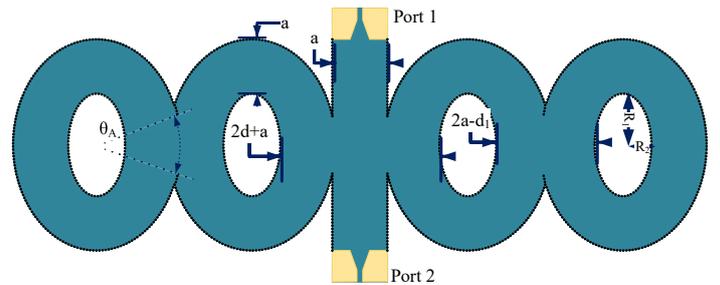

*Fig. 15. Configuration of a wave-trapper with four feedback loops (a=0.7 in, b=32 mil, $\varepsilon_r$=3.38, $R_1$=0.66 in, $R_2$=0.38 in, d=0.56 in, $d_1$=0.15 in, $\theta_A$=34° and $L_A$=0.1 in).*

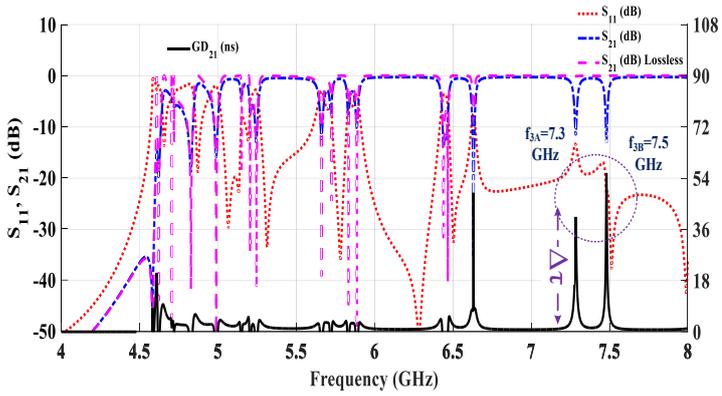

*Fig. 16. Scattering parameters and the group delay of the wave-trapper of Fig. 15.*

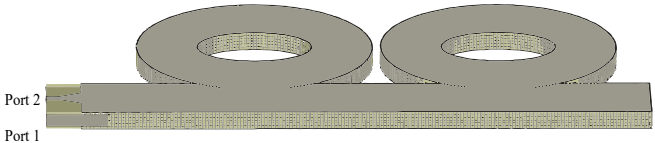

*Fig. 17. Cascading 12 cells (6 layers and two cells in each layer) of Fig. 10 configuration.*

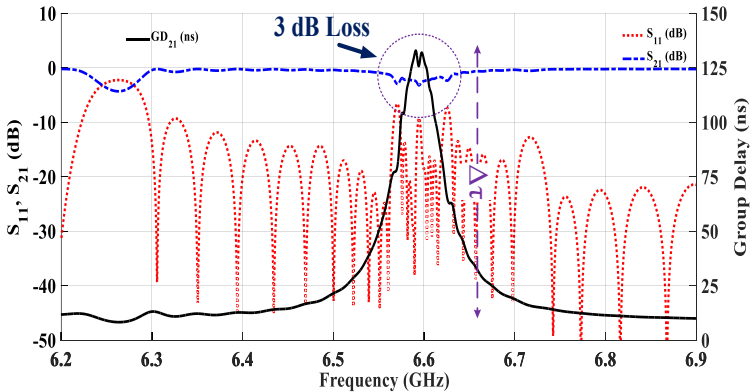

*Fig. 18. Scattering parameters and the group delay of the wave-trapper of Fig. 17, the low loss substrate ($\tan \delta = 10^{-4}$) creates 3 dB insertion loss (conductor loss has not been considered).*

Finally, the proposed SIW based wave trapper would be compared to the only similar feedback-based DDS, i.e., the coupled-ring-line configuration [6], which its implementation has been done by using HMSIW TLs [14]. As stated before, because of using feedback mechanism, the losses are expected to be large in both structures. However, due to the different operating frequencies of these structures and using amplifier, it is not fair to compare their insertion losses, directly. Nevertheless, it can be stated that in comparison with unshielded HMSIW phaser [14], the radiation loss issue is significantly solved by our proposed SIW wave trapper. In cascaded configuration, adding amplifiers compensates the losses, and also increases the group delay swing. Cascading four active HMSIW phasers proposed in [14] results in only *4 ns* group delay swing at *60.5 GHz*. In comparison, cascading twelve passive SIW DDSs proposed in this paper, creates *120 ns* group delay swing at *6.6 GHz*.

Generally, the size and the loss of the structure still is an issue compared to some other passive DDSs [11, 13]. Also, the bandwidth of the proposed structure is limited due to periodicity. Using SIW based metamaterial transmission lines in the proposed wave trapper, could be a solution for the periodicity and the size issues [12].

It is important to note that the proposed SIW based wave trapper can be used as the building block for different DDSs with arbitrary phase responses and group delays. Due to similarity of the group delay of the proposed wave trapper and the C-section, cascading approach can be used to engineer the resulted group delay of the cascaded wave trappers with linear, quadratic or semi-arbitrary phase responses [4, 5].

## 5. Experimental Results

To verify the simulation results, the modified wave trapper with a single loop (Fig .10) was implemented on Rogers RO4003 substrate (Fig .19). Scattering parameters and the group delay of the device were measured from *5 GHz* to *8 GHz*, as shown in Fig. 20. Comparing these results to the simulation data shows good agreement between them. From experimental results, it can be seen that the maximum of insertion loss (at the group delay peaks) is about *7 dB*, which is *3 dB* higher than the simulation one. The group delay peaks are *2 ns* less than the results shown in Fig. 11. Also, a *100 MHz* frequency shift can be seen from experimental results.

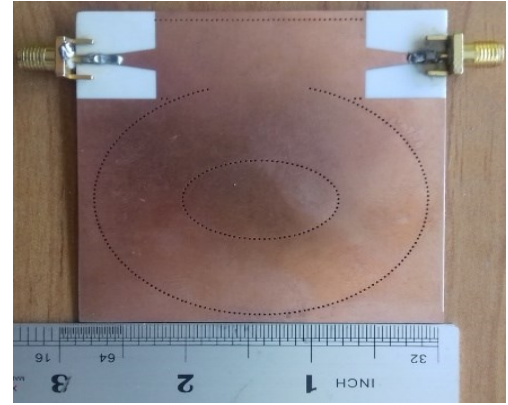

*Fig. 19. The fabricated prototype of the wave trapper with improved γ-junctions.*

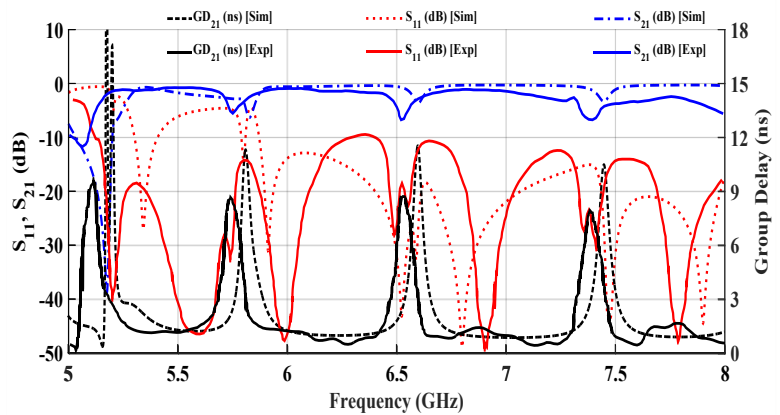

*Fig. 20. Scattering parameters and the group delay of the wave-trapper of Fig. 19 (dashed lines are simulation results).*

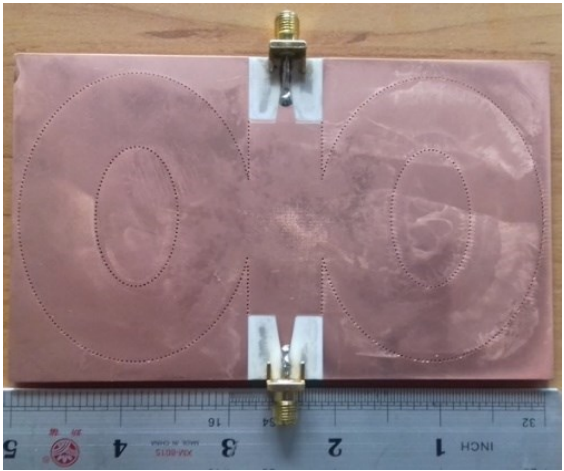

*Fig. 21. The fabricated prototype of the wave trapper with two feedback loops.*

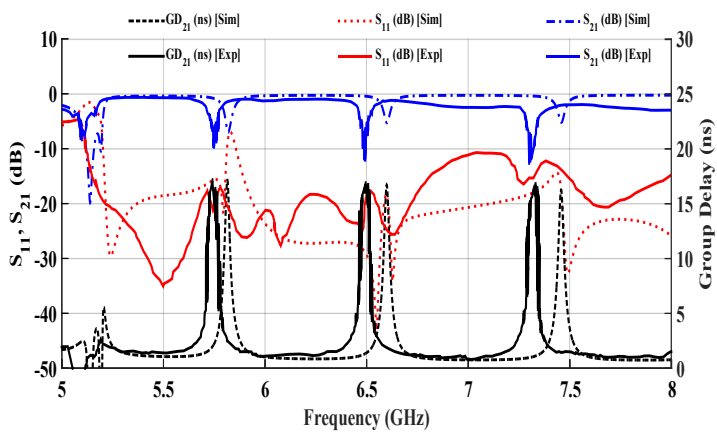

*Fig. 22. Scattering parameters and the group delay of the wave-trapper of Fig. 21 (dashed lines are simulation results).*

The second prototype is a wave trapper with two feedback loops (Fig. 21). As mentioned in the previous section, adding more feedback loops to the configuration, increases the group delay peaks. It can be seen from the group delay results illustrated in Fig. 22, where the group delay peaks are about *13 ns*. This value is *3 ns* less than the simulated group delay response. With an increase in the group delay peaks, insertion loss will also increase, which results in about *10 dB* loss at peak frequencies (Fig. 22). This loss is *4 dB* more than the estimated value.

## 6. Conclusion

The wave trapping technique described above, made it possible to create a DDS structure. The proposed configuration only consists of transmission lines and power dividers (combiners), which adds simplicity to the proposed design. The formulation for the group delay was done by following the input wave through the structure. The formula shows a direct relationship between the group delay peak value and injected energy into the feedback loop. In the following, implementation was done by using SIW TLs, which not only are dispersive, but they can also be compacted in a multi-layer configuration. By using these traits and a low loss substrate, a high group delay of *120 ns* could be reached at *6.6 GHz*. Injecting more power into the feedback loop, which can be done by improving power dividers or by adding an extra loop to the structure, results in higher group delay peaks (*8 ns* and *13 ns*, respectively).

Improving the wave trapper to increase the group delay will be based on: decreasing SIW loss by using AFSIW, replacing SIW by other transmission lines, and even by changing the configuration. Different complicated configurations could be proposed based on this simple structure, and each one of them could be realized and improved in many ways. Moreover, the proposed structures can form the fundamental building blocks for engineering more complex dispersion profiles by cascading mechanism.